\documentclass[fleqn,12pt,twoside]{article}
\usepackage[sort&compress,square,comma]{natbib}
\usepackage{espcrc1}
\usepackage{amsmath,amsfonts,amssymb}
\usepackage{graphicx}

\title{Memory and Kovacs effects in the parking-lot model: an
approximate statistical-mechanical treatment}
\author{G. Tarjus\address[LPTL]{Laboratoire de Physique
Th\'eorique des Liquides, Universit\'e Pierre et Marie Curie, 4, place
Jussieu, 75252 Paris Cedex, 05 France} and P. Viot\addressmark[LPTL] 
}

\begin{document}
\maketitle

\begin{abstract}
The parking-lot  model provides a qualitative  description of the main
features of the phenomenology  of granular compaction.  We derive here
approximate kinetic equations for this model, equations that are based
on   a $2-$parameter  generalization   of  the  statistical-mechanical
formalism first proposed  by  Edwards  and  coworkers.  We  show  that
history-dependent effects,   such as memory   and  Kovacs effects, are
captured by this approach.
\end{abstract}
\section{Introduction}
The term of ``glassy-dynamics''    is now commonly used  to   describe
out-of-equilibrium systems that  display such generic features as very
slow kinetics that prevent the system from reaching equilibrium in any
reasonable  experimental timescale, aging phenomena, history-dependent
processes like hysteresis and memory  effects.  Among such systems are
the       ``not-too-strongly''             vibrated           granular
materials\cite{KFLJN95,NKBJN98,BKNJN98,JTMJ00,PB02}.   In  the  recent
years,  there  has   been   a  surge of    research  activity  in  this
field\cite{BCKM98,B02,LN01},  partly  driven   by  the    goal  of  providing    a
statistical-mechanical   description  of   these    out-of-equilibrium
situations\cite{EO89,ME89,E94,EG99}\cite{BKLS00,BCL02,DL02b,L02,BFS02,CFN03,PrB02}.

In this  note,   we consider  an   approximate  statistical-mechanical
description of   the  parking-lot-model  (PLM) for  vibrated  granular
materials\cite{JTT94,KB94,NKBJN98,BKNJN98,KNT99,TTV99,TTV01}  that  is
based     on  the    formalism      proposed     by   Edwards      and
coworkers\cite{EO89,ME89,E94,EG99}.    Despite  its  simplicity,   the
one-dimensional  model   of   random  adsorption-desorption    of hard
particles  (PLM) has the merit  of  being  a microscopic,  off-lattice
model that mimics many features of the compaction of a vibrated column
of grains.  It also has, we hope, a didactic value as to the nature of
several canonical characteristics of ``glassy  dynamics''. Many of the
properties of the  model, that can  be obtained either analytically or
by  computer   simulation,   have  been   already  described    in the
literature\cite{JTT94,KB94,NKBJN98,BKNJN98,KNT99,TTV99,TTV00,TTV01,TTV03}. Our
main focus here is on  memory effects, including the so-called  Kovacs
effect    first   observed    in  glassy    polymers\cite{K63,S78}(see
also\cite{BB02,MS03,BBDG03,B03}),  and  on  the  ingredients  that are
needed  in a statistical-mechanical   description to account for  such
effects.

\section{The model and its properties} 

The parking-lot model is one-dimensional process in which hard rods of
length $\sigma$ are deposited at random positions on a line at rate $k_{+}$
and are  inserted   successfully only  if they  do   not overlap  with
previously adsorbed particles; otherwise  they are rejected. Moreover,
all deposited particles can desorb, i.e., be removed  from the line at
random with a rate $k_{-}$.  For convenience, the  unit time is set to
$1/k_+$, and the unit length to  $\sigma$.  With this  choice of units, the
only control  parameter in  the  model  is $K  =  k_{+}/k_{-}$.   When
desorption is forbidden ($k_-=0$), the model corresponds to the purely
irreversible   one-dimensional  random   sequential adsorption   (RSA)
process\cite{E93,TTVV00}, also known  as the car parking problem,  and
all the properties of the system as a function of time can be obtained
exactly. Connection to  the compaction of a  vibrated column of grains
is made  by considering the line as  average  layer (a $2$-dimensional
model would be more realistic,  but the qualitative behavior would not
be   altered), the time   as  the number of   taps, and  $1/K$ as the
tapping strength that controls the fraction  of particles ejected from
the layer at each tap.

When $1/K$  is not strictly  equal to zero,  adsorption and desorption
are competing  mechanisms that  drive the  system  to a  steady  state
corresponding to   an equilibrium fluid  of  hard  rods at a  constant
activity $1/K$.  All the  properties of the  steady-state can  also be
obtained exactly.

The compaction  kinetics of the parking-lot  model at constant  $K$ is
described by
\begin{equation}\label{eq:1}
\left.\frac{\partial\rho }{\partial t}\right|_K=\Phi(t)-\frac{\rho(t)}{K}
\end{equation}
where $\rho(t)$ the density of hard rods  on the line at  time $t$ and $\Phi(t)$ is  the fraction of the line  that is available  at  time $t$ for
inserting    a new particle,  i.e.,  the   probability associated with
finding  an interval free of particles  (a ``gap'') of length at least
$1$.  The quantities  $\rho$ and $\Phi$  can be obtained from the
$1-$gap distribution function $G(h,t)$ which is the density of gaps of
length $h$ at time $t$ via a number of ``sum rules'':
\begin{align}
\rho(t)&=\int_0^\infty dh G(h,t)=1-\int_0^\infty dh hG(h,t),\label{eq:2}\\
\Phi(t)&=\int_1^\infty dh(h-1)G(h,t).\label{eq:3}
\end{align}
The $1-$gap distribution function   $G(h,t)$ obeys a  kinetic equation
that  also involves   $2-$gap distribution  function.  Similarly,  the
kinetic equation  of   the $2$-gap distribution   involves the $3-$gap
distribution function, and  so on\cite{TTV00}.  The resulting infinite
hierarchy    of  coupled equations   cannot     in general  be  solved
analytically (exceptions are the  RSA,  when $k_-=0$ and   equilibrium
when $t\to+\infty$).
\begin{figure}[t]
\centering
\resizebox{8cm}{!}{\includegraphics{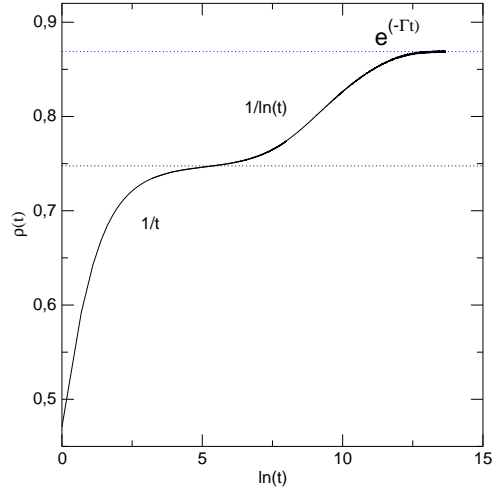}}
\caption{Logarithmic-linear plot of the density versus time for the 
parking-lot    model     at   a   constant       tapping      strength
$1/K=0.0002$}\label{fig:1}
\end{figure}

The  description of the  kinetics of  the parking lot  model at large,
constant  $K$  (i.e.   small,   constant  tapping  strength)   can  be
summarized as follows: the density  increases monotonically during the
process, and the  kinetics can be considered as  a succession of  four
different  regimes (see Fig.~\ref{fig:1}):  during  a first stage, the
density increases rapidly  until a value of around  $0.65$, and it  is
followed by  an   algebraic approach toward the   saturation  density,
$\rho_{JL}=0.747\ldots$  of the model   without desorption (RSA);  around this
density there is a crossover to  a still slower $1/ \ln(t)$ compaction
regime that is reminiscent  of what is  exactly observed when $1/K\to0$,
and  finally there  is  an  exponential  approach to  the steady-state
(equilibrium) density with a rate $\Gamma\sim \frac{(\ln(K))^3}{K^2}$. For the
two first regimes, desorption has a negligible effect and only the two
last regimes are  relevant  for  the compaction kinetics  of  vibrated
granular  materials   (although  the approach  to   equilibrium may be
prohibited when $K$ is very large).

\begin{figure}[t]
\centering
\resizebox{8cm}{!}{\includegraphics{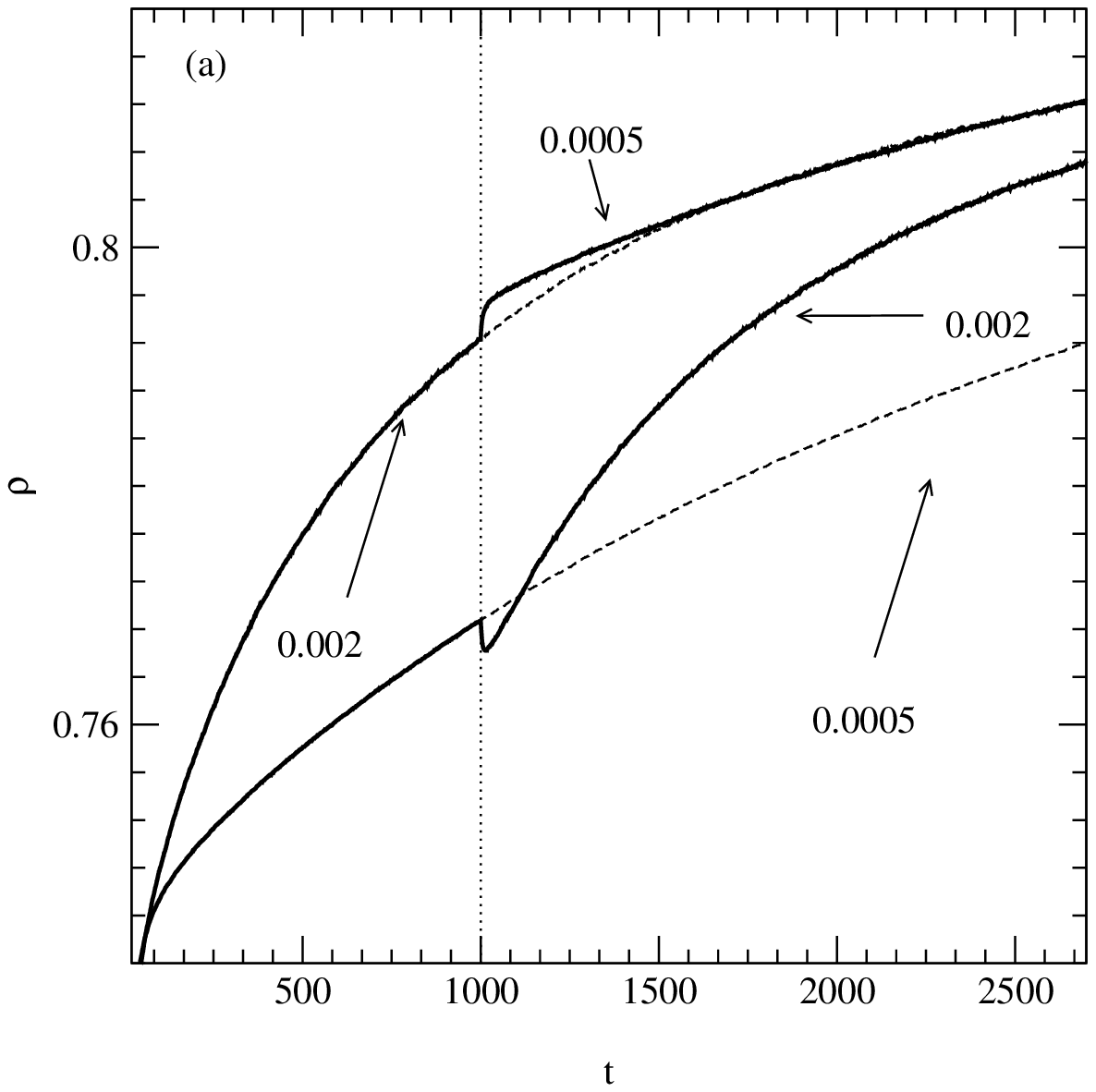}}\resizebox{8cm}{!}{\includegraphics{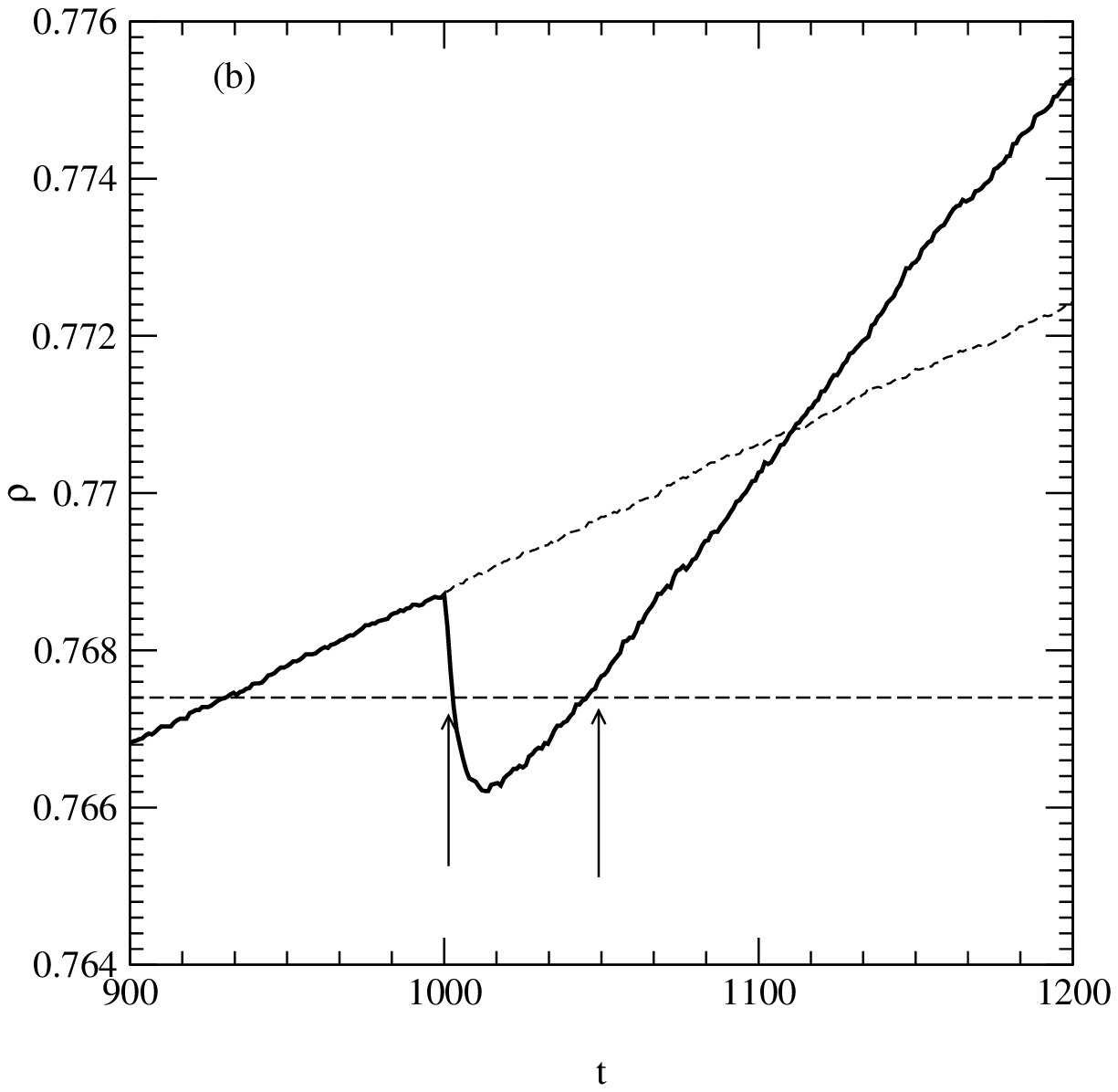}}
\caption{Memory effect: (a) density versus time when the tapping stength
is  switched from   $5.e^{-4}$  to $2.e^{-3}$  (lower  full curve)  at
$t=1000$ and from  $2.e^{-3}$ to  $5.e^{-4}$  (upper full curve).  The  dashed
curves correspond  to a  constant  strength, $5.e^{-4}$ (lower curve)  and
$2.e^{-3}$ (upper  curve); (b) Zoom up on  the region around $t=1000$ when
one  switches from $1/K=5.e^{-4}$ to  $2.e^{-3}$. The arrows denote two points
at the same density and same tapping strength $1/K$.}\label{fig:2}
\end{figure}

Experimentally, it has been observed that granular compaction exhibits
history-dependent   phenomena.     In    particular,  Josserand   {\it
et.  al}\cite{JTMJ00}  have shown that  memory   effects are seen when
changing the tapping  strength during the compaction kinetics. Similar
effects have been  found in the   parking-lot model, when  the tapping
strength $1/K$ is  switched at a given  time to a  larger (or smaller)
value:  see Fig.~\ref{fig:2}

The obvious lesson  that one  can   draw such memory effects  is  that
whereas equilibrium is fully described by one thermodynamic parameter,
the  density $\rho$, out-of-equilibrium   situations require at least one
additional ``thermodynamic''  parameter.  This parameter cannot simply
be the tapping strength  since the system can  be found in states
characterized by the same tapping strength  $1/K$ and the same density
$\rho$, that nonetheless  evolve in different  ways under further tapping
with the same  strength $1/K$ :  this is illustrated in  Fig.~\ref{fig:2}b. A
natural candidate for an  additional thermodynamic parameter is  the
available line fraction $\Phi$. One can indeed  check that the two states
described above do correspond to distinct values of $\Phi$.

\section{Statistical-mechanical formalism with two thermodynamic parameters}

Following the ideas put forward by Edwards and
coworkers\cite{EO89,ME89,E94}, we consider a statistical-mechanical
description of the system in which all possible microstates
characterized by a (small) number of fixed macroscopic quantities are
assumed to be equiprobable (``flat'' or microcanonical
distribution). In addition to fixing the density $\rho$, the parameter
originally selected in the compactivity-based description of granular
media by Edwards and coworkers, we also constrain the microcanonical
ensemble by fixing the available line fraction $\Phi$ to account for the
above discussion. 

Denoting by $A$ the total length available for insertion of a
particle center ($A=\Phi L$), the configurational integral with the
constraints of fixed $A$, fixed system size $L$, and fixed number of
particles $N$ is obtained as\cite{TV03}

\begin{equation}\label{eq:4}
Z(L,N,A)=\int_0^L\ldots \int_0^L\prod_{i=1}^Ndh_i\delta\left(L-N-\sum_{i=1}^{N}h_i\right)\delta\left(A-\sum_{i=1}^{N}\theta(h_i-1)(h_i-1)\right),
\end{equation}
which can be rewritten  as
\begin{equation}
Z(L,N,A)=\int_Cdz\int_{C'}dy
\exp\left(L\left[z(1-\rho)+y\Phi+\rho
\ln\left(\frac{z+y(1-\exp(-z))}{z(z+y)}\right) \right]\right)
\end{equation}
where $C$ and $C'$ denote two  closed contours. In the macroscopic
limit, $N\to\infty$, $L\to\infty$, $A\to\infty$ with $\rho$ and $\Phi$ fixed, one can  use a
saddle-point method to evaluate the integrals, which leads to 
\begin{equation}
Z(L,N,A)\simeq \exp(Ls(\rho,\Phi))
\end{equation}
where the entropy density $s(\rho,\Phi)$ is given by
\begin{equation}
s(\rho,\Phi   )=(1-\rho)z+y\Phi +\rho  \ln\left(\frac{z+y(1-\exp(-z))}{z(z+y)}\right),
\end{equation}
with $z\equiv z(\rho,\Phi)$ and $y\equiv y(\rho,\Phi)$ solutions of the two coupled
equations
\begin{align}
\left(\frac{1-\rho }{\rho}\right)&=\frac{1}{z}+\frac{1}{z+y}-\frac{1+ye^{-z}}
{z+y(1-e^{-z})},\label{eq:5}\\
\frac{\Phi}{\rho}&=\frac{1}{z+y}-\frac{1-e^{-z}}
{z+y(1-e^{-z})}.\label{eq:6}
\end{align}

The  gap  distribution functions    can  also derived along  the  same
lines\cite{TV03}, which leads to
\begin{align}\label{eq:7}
G_{Ed}(h;\rho)=\begin{cases}
\rho\displaystyle\frac{z(z+y)}{z+y\left(1-e^{-z}\right)}e^{-zh}         &
\mbox{for                           }                           h<1,\\
\rho\displaystyle\frac{z(z+y)}{z+y\left(1-e^{-z}\right)}e^{-(zh+y(h-1))}
& \mbox{for } h>1.  \end{cases}
\end{align}
It can also be shown that the multi-gap distribution functions satisfy
a  factorization   property,   e.g., $G_{Ed}(h,h';\rho,\Phi   )=G_{Ed}(h;\rho,\Phi
)G_{Ed}(h';\rho,\Phi )$. Note that the  $1$-gap distribution function is a
piecewise    continuous function that      obeys the exact  sum rules,
Eqs.~(\ref{eq:2}) and ~(\ref{eq:3}). 

A detailed comparison between the predictions of this
statistical-mechanical treatment and simulation data {\it at the same
$\rho$ and $\mathit \Phi$} can be found in Ref.\cite{TV03}. The conclusion is that
although not exact, and even missing some qualitative features in the
limiting case of a purely irreversible RSA process, the approach
provides an overall good description of the data. (It becomes of
course exact in the steady state since this latter corresponds to an
equilibrium situation.)

\section{Approximate description of the compaction kinetics}

A quasi-thermodynamic approach is useful because  it allows to predict
the structure  of the system   (correlation functions) as well  as the
fluctuations  at any given state  point (here assumed to characterized
by $\rho$ and $\Phi$).  In  the cases where  phase transitions occur, it can
also  provide   interesting  constraints  and relations   between  the
parameters characterizing  the phases at   the transition  or help  to
determine the limit of  stability of the  phases in mean-field like
treatments.  However, one still   faces the problem of predicting  the
state of the system {\it for a given protocol and a given history}: in
the simplest case, for   a given time  $t$ and  a given tapping   strength
$1/K$. 

We    thus  want  to     push    the Edwards  formalism    one   step
further.  Predicting the trajectory made by  the system in the $(\rho,\Phi)$
diagram for a   given  protocol amounts   to  determine a   system  of
equations relating $\rho$  and $\Phi$  with $t$ and   $K$. 
 Eq.~(\ref{eq:1}) is one such equation. To obtain another equation for
the evolution of $\Phi$, we start for the exact kinetic equation for the
$1-$gap distribution function for $h\geq 1$:
\begin{equation}
\left.\frac{\partial G(h,t)}{\partial t}\right|_K=-(h-1)G(h,t)+2\int_{h+1}^\infty dh'G(h',t)
-\frac{2G(h,t)}{K}+\frac{1}{K}\int_0^{h-1}dh'G(h',h-1-h',t),\label{eq:8}
\end{equation}
where $G(h,h',t)$ is the $2$-gap distribution function that satisfies
the ``sum rule''
\begin{equation}\label{eq:9}
\int_0^\infty dh'G(h,h',t)=\rho(t)G(h,t).
\end{equation}
Note that the approximate $2-$gap distribution function $G_{Ed}(h,h')$
satisfies the above sum rule.
By multiplying Eq.~(\ref{eq:8}) by $(h-1)$, integrating over $h$
from $1$ to $\infty$, and using the sum rules,
Eqs.~(\ref{eq:2}),~(\ref{eq:3}) and (\ref{eq:9}), one obtains an exact
equation for the evolution of $\Phi$:
\begin{equation}\label{eq:10}
\left.\frac{\partial\Phi (t)}{\partial t}\right|_K=\frac{2}{K}(1-\rho(t)-\Phi(t))
-\int_1^\infty dh (h-1)^2G(h,t) +\int_2^\infty dh(h-2)^2G(h,t).
\end{equation}
If one now inserts   the  approximate expression  of  the  $1$-gap   distribution
function,  Eq.~(\ref{eq:7}), in the above equation, one gets
\begin{equation}\label{eq:11}
\left.\frac{\partial\Phi  }{\partial t}\right|_K=\frac{2(1-\rho(t)-\Phi(t))}{K}-2\Phi(t) \frac{1-e^{-(y(t)+z(t))}}{y(t)+z(t)}.
\end{equation}

Eqs. (\ref{eq:1}),~(\ref{eq:5}),~(\ref{eq:6}) and (\ref{eq:11}) form a
closed set of equations whose solution for given initial conditions
completely characterizes the system and its evolution.

We  first    test  the  accuracy  of   the  above  approximate kinetic
description  in two limiting  cases  for which  the exact solution  is
known: the purely irreversible RSA case  ($1/K=0$) and the approach to
the steady  state  ($t\to+\infty$) for a given  rate  $K$.  The  steady state
corresponds to  the solution $z_\infty =\frac{\rho_\infty}{1-\rho_\infty}$,  $y_\infty=0$, where
$\rho_\infty$  is  the equilibrium density of  hard  rods at constant activity
$1/K$:  $\rho_\infty=K\Phi_\infty$     with    $\Phi_\infty=(1-\rho_\infty)\exp\left(\frac{-\rho_\infty}{1-\rho_\infty}
\right)$; it is thus the exact result. The approach to the steady state
is   obtained   by linearizing  the    kinetic   equations around  the
equilibrium  solution. One then  obtains  the following coupled linear
differential equations :
\begin{align}
\left.\frac{\partial y(t)}{\partial t}\right|_K&=-2\frac{(z_\infty+1)(e^{-2z_\infty}-1) 
+2(1+z_\infty^2)e^{-2z_\infty} }{z_\infty
(2e^{z_\infty }-z_\infty^2-2z_\infty -2)}y(t)\\
\left.\frac{dz(t)}{dt}\right|_K&=-\frac{6
e^{z_\infty} -2z_\infty ^2-6-8z_\infty +(2+z_\infty^2 )z_\infty e^{-z_\infty }}
{z_\infty (2e^z_\infty-z_\infty^2-2z_\infty -2)}y(t)-\frac{z_\infty +1}{z_\infty }(z(t)-z_\infty).
\end{align}

The associated eigenvalues are  all negative and that corresponding to
the inverse of relaxation time goes for large $K$ as
\begin{equation}
\tau^{-1}=\Gamma \sim  \frac{\ln(K)^2}{3K}.
\end{equation}
The relaxation time is of the same order of magnitude as that obtained
in  the  simple adiabatic  approximation\cite{BKNJN98}   and  is  much
smaller than the  exact result (see  above). The error comes  from the
inability  of    the Edwards  approximation      to account  for   the
non-exponential behavior of  the $1-$gap distribution function for $0\leq
h\leq1$.

One the other hand the approximate kinetic description is very good in
the purely irreversible case, $1/K=0$, (but it still misses some
features, as discussed above and in Ref.\cite{TV03}). Then Eqs.~(\ref{eq:1}) and
(\ref{eq:11}) simplify to

\begin{figure}[t]
\centering 

\resizebox{8cm}{!}{\includegraphics{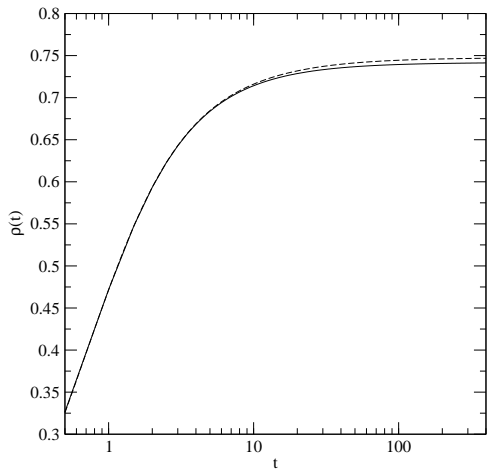}}
\caption{Density  versus time for the model without desorption (RSA): the
full curve corresponds to the  approximation, the dashed curve
is the exact result.}\label{fig:3}
\end{figure}
\begin{align}
\frac{\partial\rho }{\partial t}&=\Phi(t)\label{eq:12}\\\label{eq:13}
\frac{\partial\Phi  }{\partial t}&=-2\Phi(t) \frac{1-e^{-(y(t)+z(t))}}{y(t)+z(t)},
\end{align}
where $y$ and $z$ are related to  $\rho$ and $\Phi$ by Eqs.~(\ref{eq:5}) and
(\ref{eq:6}). It is easy to  show that at long times  $z(t)$ goes to a
finite  limit whereas $y(t)$ goes   to $\infty $   as $t$. As  a result the
kinetics  approaches  a non-trivial  jamming   limit with an algebraic
$1/t$  behavior.  The numerical solution   of the approximate equation
shows  that the saturation density at  the jamming limit $\rho_{JL}^{Ed}\simeq
0.7422$ is  very  close to the exact   value, $\rho_{JL}= 0.74759\ldots$.  The
overall  agreement with the exact  density evolution  is very good, as
shown in Fig.~\ref{fig:3}.

\begin{figure}[t]
\centering 

\resizebox{8cm}{!}{\includegraphics{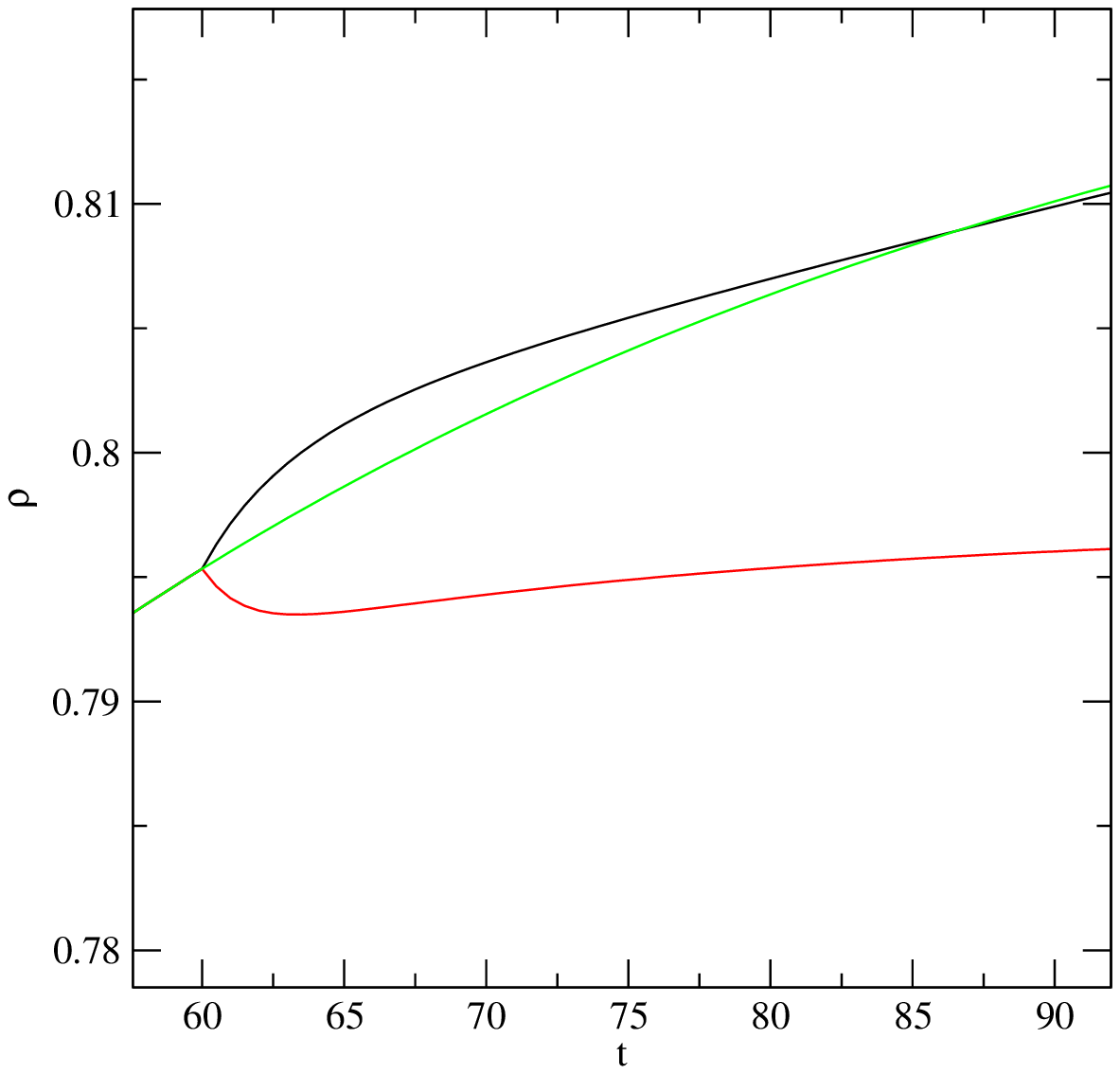}}\resizebox{8cm}{!}{\includegraphics{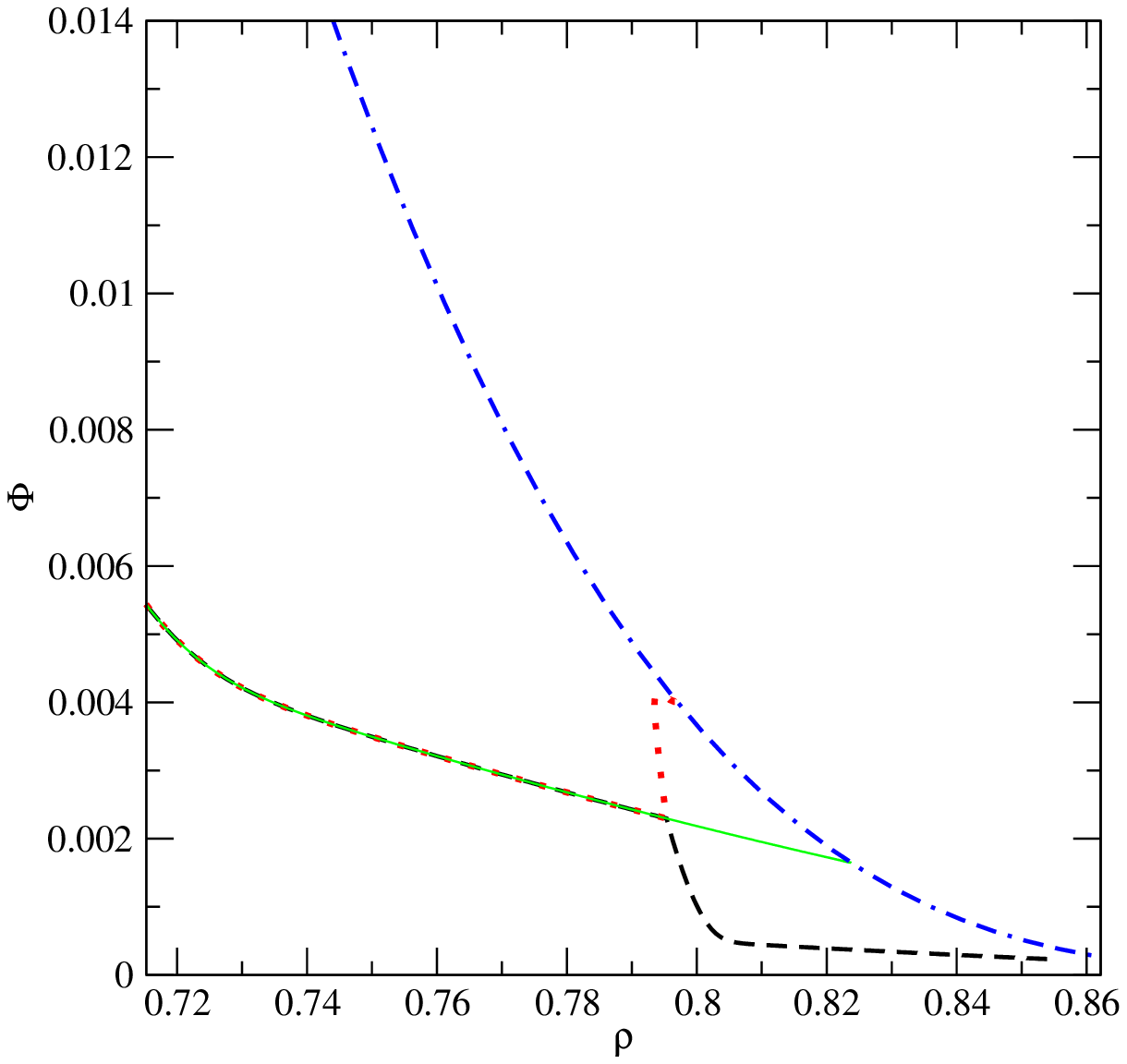}}
\caption{(a) Approximate time evolution of the density  for three cases:
$K=500$ (full  curve), $K=500$ from  $t=0$ to  $t=60$ and  $K=200$ for
$t>60$ (dotted curve)  and $K=500$ from  $t=0$ to $t=60$ and  $K=5000$
for $t>60$ (dashed curve). (b)Parametric plot of insertion probability
$\Phi$ versus density  for the three  above cases.  For completeness, the
equilibrium    insertion     probability    is   added       (dash-dot
curve). }\label{fig:4}
\end{figure}

\begin{figure}[ht]
\centering
\resizebox{8cm}{!}{\includegraphics{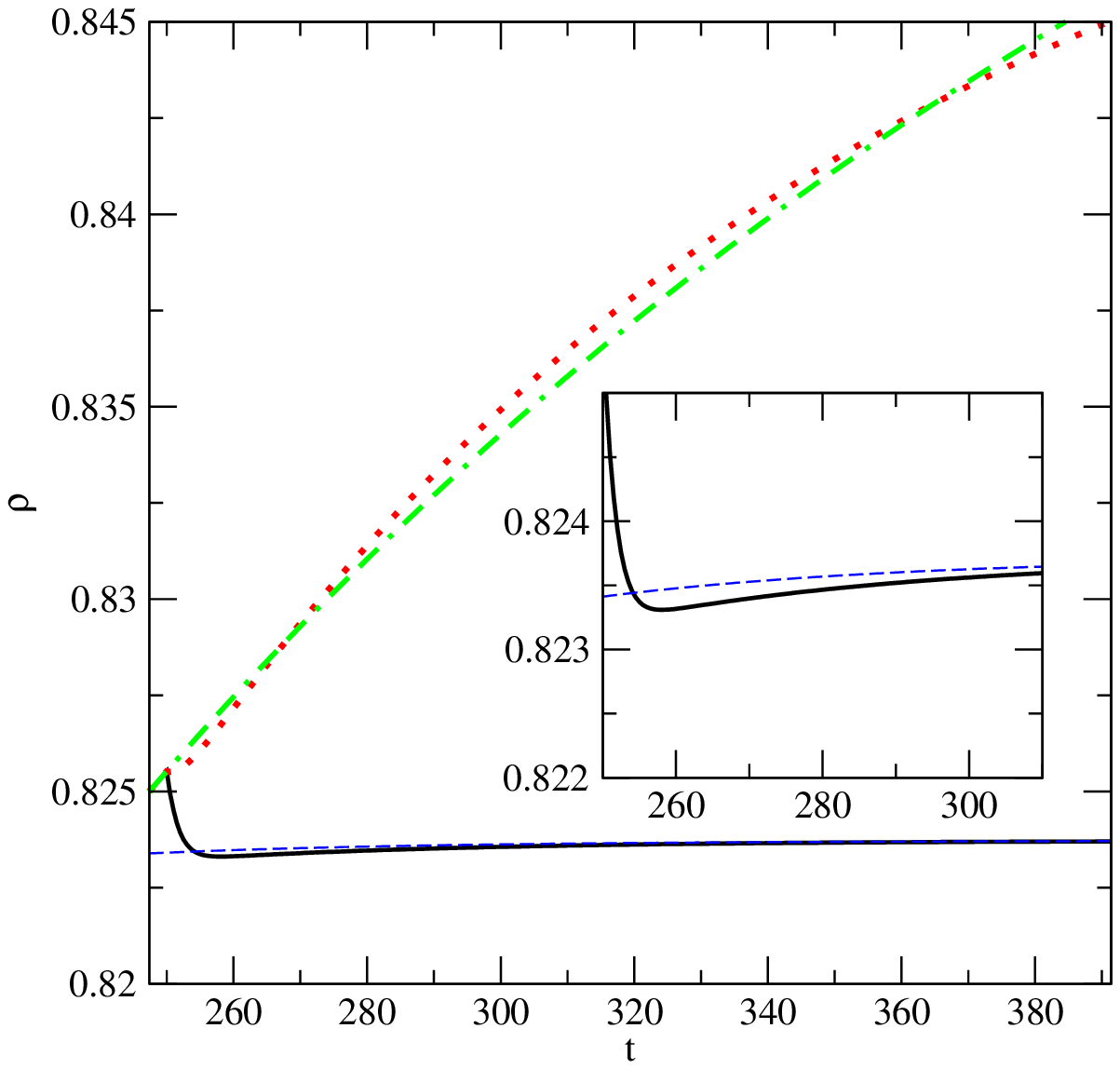}}\resizebox{8cm}{!}{\includegraphics{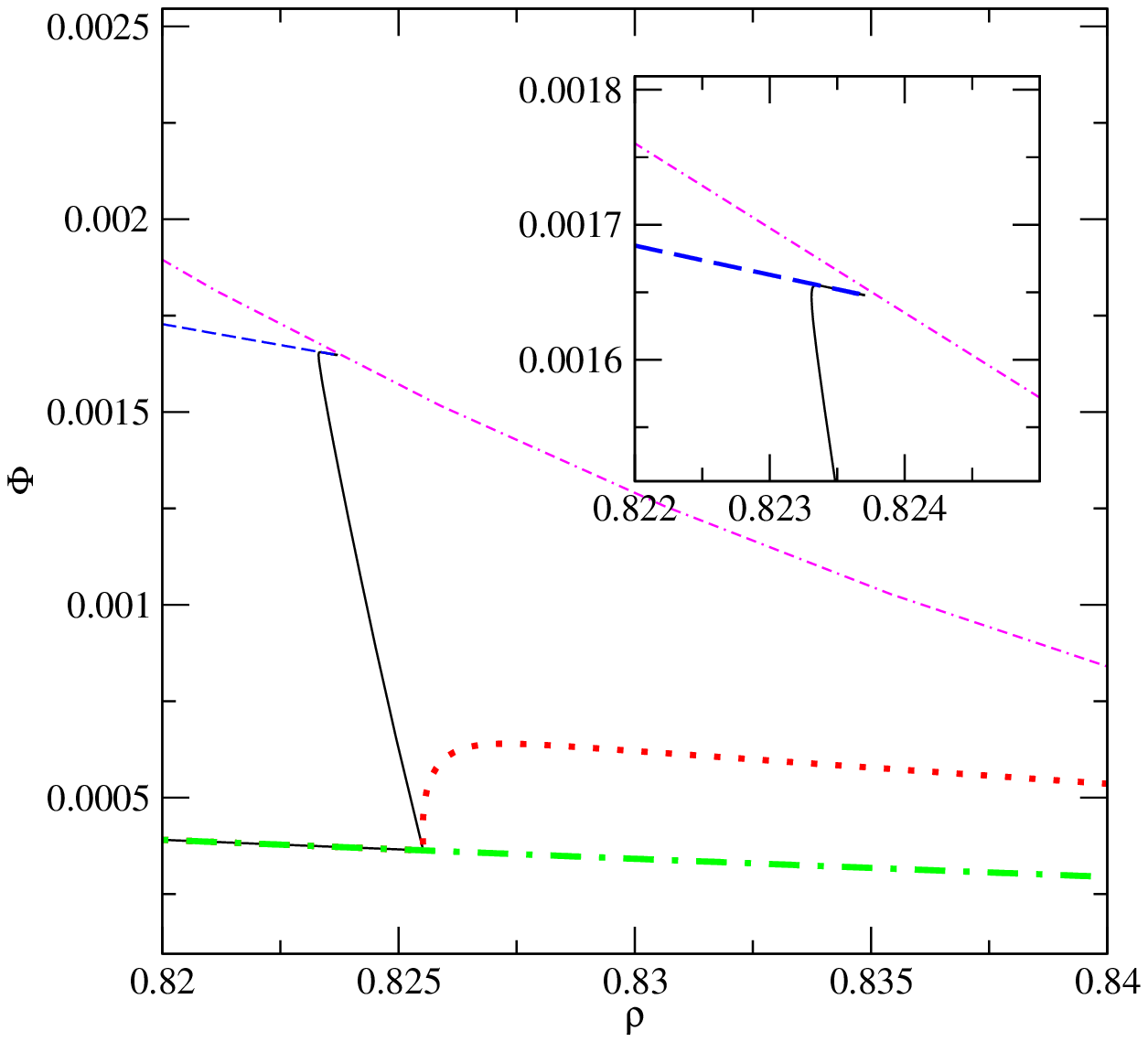}}
\caption{(a)Approximate time evolution of the density   for three
cases:$K=500$  (lower  curve),  $K=5000$ (dot-dashed curve),  $K=5000$
from  $t=0$  to $t=250$  and $K=500$   for  $t>250$ (full   curve) and
$K=5000$   from $t=0$  to  $t=250$ and   $K=2000$   for $t>60$ (dotted
curve). (b)Parametric  plot  of the  insertion probability $\Phi$  versus
 density for the four above cases:  For completeness,the equilibrium
insertion probability is added (dashed curve).}\label{fig:5}
\end{figure}

\section{Memory and Kovacs Effects}
Finally, we apply the approximate kinetic equations to the description
of memory  effects.  By introducing $\Phi$ as  the second state variable,
one expects to  obtain a response to a   sudden change of  the tapping
strength  that  captures  these memory     effects.   This is   indeed
illustrated in  Fig.~\ref{fig:4}a  that shows how  the density evolves
when  $K$ is  changed   at $t=60$ from   $K=500$ to  $K=5000$ and from
$K=500$ to   $K=200$.  At longer  times, note  that the  density curve
corresponding to the change from $K=500$ to  $K=200$ crosses two times
the density curve  calculated when $K=200$ is  kept constant along the
process.  This is the signature of a Kovacs-like  effect which is more
clearly displayed in the parametric plot of Fig.~\ref{fig:4}b: instead
of going monotonically to the final value, the density decreases first
and  passes by a  minimum  after increasing again. This  non-monotonic
behavior   of the  density (or   the volume)  is  precisely the Kovacs
effect. The same phenomenon is illustrated for different values of the
tapping strength  in  Figs~\ref{fig:5}a and  b: $K$ is   switched from
$5000$ to $500$ or from $5000$ to $2000$  after a time $t=250$. In the
former  case, the density  reached at $t=250$  when $K$  is changed is
higher   than  the  final  (equilibrium) density    reached at $K=500$;
nonetheless,  the density starts by first  decreasing  to a value less
than the    final density  until  it  reaches   a minimum and  finally
increases  again.  On the other  hand,  no minimum  is observed in the
latter case.

\begin{figure}[ht]
\centering
\resizebox{8cm}{!}{\includegraphics{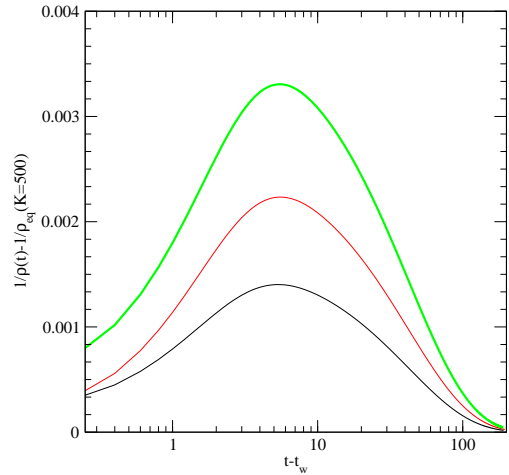}}
\caption{Kovacs effect in the approximate description of the
parking-lot model : $1/ \rho(t)-1/ \rho_{eq}(K=500)$  versus $t-t_w$. In the
upper curve $K$ is switched from  $K=5000$ to $K=500$ at $t_w=240$, in
the intermediate  curve $K$  is  switched from $K=2000$ to  $K=500$ at
$t_w=169$, and in the  lower  curve, $K$ is  switched from  $1000$  to
$500$ at $t_w=139$.}\label{fig:6}
\end{figure}

The  Kovacs effect is traditionally represented  as a  ``hump'' in the
volume  as a function  of time\cite{K63,BBDG03,MS03}. The position and
the height of the   hump vary with the   waiting time, i.e.,  the time
spent at   the initial tapping strength   (recall that the  process of
compaction is always  out of equilibrium), and  with the  amplitude of
the shift  in the tapping  strength  (or in other  glassy systems, the
shift in temperature).   The   same behavior  is  found  here.  It  is
illustrated  in Fig.~\ref{fig:6} where we  show  the evolution of  the
inverse  density for  different   protocols to reach  the  equilibrium
steady  state at $K=500$:  in  the upper curve,   $K$ is switched from
$5000$  to   $500$ at $t_w=240$,  in  the  intermediate  curve  $K$ is
switched from  $2000$  to $500$ at  $t_w=169$ and  in the lower  curve
$K=1000$   at $t_w=139$. The  waiting  time $t_w$  is  chosen that the
density   reached at  that   time     is  equal  to final     density,
$\rho_{eq}(K=500)$. One observes that the height of the hump increases as
the amplitude of the shift increases. The Kovacs effect, also observed
in the simulation data (not shown here), results from competing trends
that are qualitatively  well accounted for  by the present description
in   terms of two thermodynamic   parameters.  Including $\Phi$ in  the
statistical-mechanical  treatment   is  crucial for  reproducing  such
behavior.

\section{Conclusion}
We have applied a two-parameter statistical-mechanical formalism
inspired by the work of Edwards and coworkers to describe the
compaction kinetics of the parking-lot model. The approximation gives
a fair description of the kinetics, although it underestimates the
relaxation time characteristic of the final approach to
equilibrium. Inclusion of a second thermodynamic parameter allows one
to qualitatively reproduce experimentally observed memory effects in
granular compaction and to predict a Kovacs effect similar to what is
observed in many glassy systems.

\end{document}